\documentclass[aps,pra,preprint,showpacs]{revtex4}
\usepackage[dvips]{graphicx}
\newcommand{\mib}[1]{\mbox{\boldmath $#1$}}

\def\C{{\bf C}}
\def\R{{\bf R}}
\def\Z{{\bf Z}}
\def\N{{\bf N}}
\def\X{{\bf X}}
\def\V{{\bf V}}
\def\x{{\bf x}}
\def\p{{\bf p}}
\def\q{{\bf q}}
\def\k{{\bf k}}
\def\w{{\bf w}}
\def\v{{\bf v}}

\def\0{{\bf 0}}

\def\im{{\rm i}}

\def\vsigma{\mib{\sigma}}

\begin{document}

\title{Dirac equation with ultraviolet cutoff
and quantum walk}

\author{Fumihito Sato}
\email[]{f.sato@phys.chuo-u.ac.jp}
\affiliation{Department of Physics,
Faculty of Science and Engineering,
Chuo University, Kasuga, Bunkyo-ku, Tokyo 112-8551, Japan}
\author{Makoto Katori} 
\email[]{katori@phys.chuo-u.ac.jp}
\affiliation{Department of Physics,
Faculty of Science and Engineering,
Chuo University, Kasuga, Bunkyo-ku, Tokyo 112-8551, Japan}

\date{14 December 2009}

\begin{abstract}
The weak convergence theorems of the one- and
two-dimensional simple quantum walks, SQW$^{(d)}, d=1,2$,
show a striking contrast to the classical counterparts,
the simple random walks, SRW$^{(d)}$.
In the SRW$^{(d)}$, distribution of position $\X(t)$ of the particle
starting from the origin converges to the Gaussian distribution
in the diffusion scaling limit, in which the
time scale $T$ and spatial scale $L$ both go to infinity 
with keeping the ratio $L/\sqrt{T}$ to be finite.
On the other hand, in the SQW$^{(d)}$, the ratio 
$L/T$ is kept to define the pseudovelocity
$\V(t)=\X(t)/t$, and then all joint moments of the components
$V_j(t), 1 \leq j \leq d$, 
of $\V(t)$ converges in the $T=L \to \infty$ limit.
The limit distributions have novel structures such that
they are inverted-bell shaped and the supports of them
are bounded.
In the present paper we claim that these properties of the
SQW$^{(d)}$ can be explained by the theory of relativistic 
quantum mechanics. 
We show that the Dirac equation with a proper ultraviolet cutoff
can provide a quantum walk model in three dimensions,
where the walker has a four-component qubit.
We clarify that the pseudovelocity $\V(t)$
of the quantum walker, which solves the Dirac equation, 
is identified with the relativistic velocity.
Since the quantum walker should be a tardyon, not a tachyon,
$|\V(t)| < c$, where $c$ is the speed of light, 
and this restriction (the causality) is the origin of
the finiteness of supports of the limit distributions
universally found in quantum walk models.
By reducing the number of components of momentum
in the Dirac equation, we obtain the limit distributions
of pseudovelocities for the lower dimensional
quantum walks. We show that the obtained limit distributions
for the one- and two-dimensional systems have the common 
features with those of SQW$^{(1)}$ and SQW$^{(2)}$.
\end{abstract}


\maketitle

\section{INTRODUCTION}

The notion of quantization of random walks 
has been widely discussed
\cite{ADZ93,Mey96,NV00,ABNVW01,TM02,Kon02,Kem03,
Amb03,Ken06,BBCKPX08}
and applications 
of quantum walk models have been studied,
for example, in information theory \cite{Gro97,NC00,Amb04}
and in solid-state physics \cite{OKAA05}.
One of the new topics in the study of quantum walks
is to discuss the relationship between quantum walk models
and the {\it relativistic quantum mechanics} 
\cite{KFK05,Str06,BES07,Str07}.
In the present paper, we concentrate on 
the {\it weak limit theorem of pseudovelocity}
of quantum walk, which was first proved by Konno
for the one-dimensional simple quantum walk
\cite{Kon02,Kon05} and then has been obtained
for other models 
\cite{GJS04,IKK04,MKK07,WKKK08,SK08,Kon09},
and connections between the solutions of the {\it Dirac equations}
and the quantum walk models are reported.

{\it Simple random walk} (SRW) is a discrete-time
stochastic process defined on a lattice
such that particle hopping is allowed only between
nearest neighbor sites at each time step.
Let $\Z$ be the set of all integers, 
$\Z=\{\dots, -2, -1, 0, 1, 2, 3, \dots\}$.
We consider the models on the $d$-dimensional
hypercubic lattices
$\Z^d=\{\x=(x_1, \dots, x_d): x_j \in \Z, 1 \leq j \leq d\}$.
For each site in $\Z^d$, there are $2d$ nearest neighbor sites
and we put $p_j$ be the probability that the particle hops to
the $j$-th nearest neighbor site, $1 \leq j \leq 2d$.
The elementary process of the SRW$^{(d)}$ is determined
by a $2d$-component vector
$\p=(p_1, \dots, p_d)$ satisfying the conditions
$0 \leq p_j \leq 1, 1 \leq j \leq 2d$, and
$\sum_{j=1}^{2d} p_j=1$.
In other words, if we prepare such a die that
it has $2d$ faces and 
the $j$-th face appears with probability
$p_j, 1 \leq j \leq 2d$, in each casting,
a path of the $d$-dimensional SRW (SRW$^{(d)}$) can be 
determined by a sequential random casting
of this die.

The $d$-dimensional {\it simple quantum walk}, SQW$^{(d)}$, 
is obtained by quantizing the SRW$^{(d)}$.
In the quantization, the die is replaced by 
a ``quantum die", which is expressed by
a $2d \times 2d$ unitary matrix $A^{(d)} \in {\rm U}(2d)$.
In the Fourier space, the nearest neighbor
hopping is also described by using a $2d \times 2d$
unitary matrix 
representing a shift operator, which is diagonal 
\begin{equation}
S^{(d)}(\k)={\rm diag} \Big[ e^{\im k_1}, e^{-\im k_1}, \dots,
e^{\im k_d}, e^{-\im k_d} \Big],
\label{eqn:Sk}
\end{equation}
where $\im=\sqrt{-1}$ and $k_{j} \in (\pi, \pi]$ denotes
the $j$-th component of the wave number vector $\k$.
The dynamics of the SQW$^{(d)}$ at each time step is then
determined by the unitary matrix
\begin{equation}
V^{(d)}(\k)=S^{(d)}(\k) A^{(d)}, \quad \k \in (-\pi, \pi]^{d}.
\label{eqn:V1}
\end{equation}
The important consequence of this quantization is
that the state of the particle (quantum walker) at each time
should be represented by a $2d$-component
vector-valued wave function, 
to which the $2d \times 2d$ unitary
matrix (\ref{eqn:V1}) is operated.
We assume that 
there are only one quantum walker at the origin $O$
at time $t=0$. 
We let the walker have an initial $2d$-component qubit 
$\q=(q_1, q_2, \cdots, q_{2d})$
with $q_j \in \C, 1 \leq j \leq 2d$ and
$\sum_{j=1}^{2d} |q_j|^2=1$.
That is, in the $\k$-space
the initial wave function is independent of $\k$
and simply given by
\begin{equation}
\widehat{\Psi}(\k, 0)= \widehat{\Psi}_0 \equiv
{^{\rm t}}(\begin{array}{cccc}
q_1 & q_2 & \cdots & q_{2d}
\end{array}),
\label{eqn:qubit}
\end{equation}
where the left-superscript ${\rm t}$ means the
transpose of the matrix.
The wave function of the walker at time 
$t \in \N_0 \equiv \{0, 1,2,3, \dots\}$ is
given in the $\k$-space and in the real space $\R^{d}$ by
\begin{equation}
\widehat{\Psi}(\k,t)=[V^{(d)}(\k)]^t \widehat{\Psi}_0
\label{eqn:Psik}
\end{equation}
and
\begin{equation}
\Psi(\x,t)=\prod_{j=1}^{d} \int_{-\pi}^{\pi} 
\frac{dk_j}{2\pi} e^{\im \k \cdot \x} 
\widehat{\Psi}(\k,t),
\label{eqn:Psix}
\end{equation}
respectively, where $\k \cdot \x=\sum_{j=1}^{d} k_j x_j$.
The probability that the quantum walker is observed
at site $\x \in \Z^{d}$ at time $t$ is given by
\begin{equation}
P(\x,t)=||\Psi(\x,t)||^2 \equiv
\Psi^{\dagger}(\x, t) \Psi(\x,t),
\label{eqn:P}
\end{equation}
where $\Psi^{\dagger}(\x,t)={^{\rm t}} \overline{\Psi}(\x,t)$
denotes the Hermitian conjugate of $\Psi(\x,t)$.
In this paper
$\bar z$ denotes the complex conjugate of $z \in \C$.
Let $\X(t)=(X_1(t), X_2(t), \dots, X_d(t))$ be
the position of the quantum walker at time $t$,
whose probability distribution is given by (\ref{eqn:P}).
The joint moment of the components $X_j(t), 1 \leq j \leq d$, 
of $\X(t)$ is then obtained by \cite{KFK05}
\begin{eqnarray}
\left\langle \prod_{j=1}^{d} (X_j(t))^{\alpha_j} \right\rangle
&\equiv& \sum_{\x \in \Z^{d}}
\prod_{j=1}^{d} x_j^{\alpha_j} P(\x,t)
\nonumber\\
&=& \prod_{j=1}^{d} \int_{-\pi}^{\pi} \frac{dk_j}{2 \pi}
\widehat{\Psi}^{\dagger}(\k,t)
\prod_{j=1}^{d} 
\left( \im \frac{\partial}{\partial k_j} \right)^{\alpha_j}
\widehat{\Psi}(\k,t)
\label{eqn:moment1}
\end{eqnarray}
for any $\alpha_j \in \N_0, 1 \leq j \leq d$.

\begin{figure}[htpb]
\begin{center}
\includegraphics[width=0.5\linewidth]{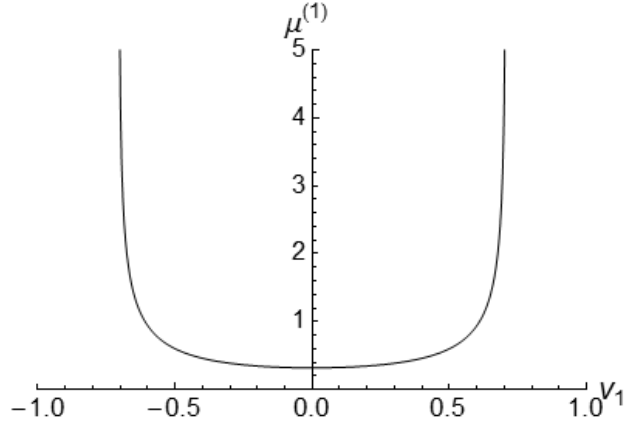}
\caption{
The one-dimensional density function
$\mu^{(1)}(v_1;|a|)$
of limit distribution of pseudovelocity
for the SQW$^{(1)}$ 
(the Konno function), when $a=1/\sqrt{2}$.}
\label{fig:fig_SatoKatori_1}
\end{center}
\end{figure}

The {\it unitarity} of the time-evolution operator (\ref{eqn:V1})
implies that in principle 
we are not able to find any
convergence property of wave function $\Psi(\x,t)$
nor of the probability distribution $P(\x,t)$
in the long-time limit $t \to \infty$ 
in the SQW$^{(d)}$.
It presents a striking contrast to the classical 
stochastic processes, SRW$^{(d)}$, which will generally converge to 
diffusion particle systems in the long-time $T \to \infty$ and
large-scale $L \to \infty$ limit 
with the diffusion scaling $L/\sqrt{T}=$ const., 
and probability laws of the diffusion particle systems 
are described by using the Gaussian distribution functions.
Konno \cite{Kon02,Kon05} discovered, however, that
in the one-dimensional model, SQW$^{(1)}$, if we consider the 
{\it pseudovelocity} defined by
\begin{equation}
V_1(t)=\frac{X_1(t)}{t},
\label{eqn:vel1}
\end{equation}
instead of the position $X_1(t)$ or
the usual velocity $dX_1(t)/dt$, 
it does converge
in a weak sense such that
any moment of $V_1(t)$ converges
to a moment of a continuous random variable as $t \to \infty$,
whose distribution is given by a novel
probability density function $\mu^{(1)}$.
That is, 
{\it Konno's weak convergence theorem} is given
as follows \cite{Kon02,Kon05}.
Consider the SQW$^{(1)}$ model driven by 
the quantum die
\begin{equation}
A^{(1)}=\left( \begin{array}{cc}
a & b \cr - \overline{b} & \overline{a} 
\end{array} \right) 
\in {\rm U}(2),
\quad a, b \in \C, \quad |a|^2+|b|^2=1, 
\label{eqn:A1}
\end{equation}
where the initial qubit of the walker is
$\q=(q_1, q_2)$.
Then for any $\alpha_1 \in \N_0$
\begin{equation}
\lim_{t \to \infty}
\Big\langle V_1(t)^{\alpha_1} \Big\rangle
=\int_{-\infty}^{\infty} dv_1 \, v_1^{\alpha_1}
\nu^{(1)}(v_1; A^{(1)}, \q),
\label{eqn:conv1}
\end{equation}
where
\begin{equation}
\nu^{(1)}(v_1; A^{(1)}, \q)=\mu^{(1)}(v_1;|a|) 
{\cal M}^{(1)}(A^{(1)}, \q)
\label{eqn:nu1}
\end{equation}
with
\begin{eqnarray}
\label{eqn:mu1}
&& \mu^{(1)}(v_1; |a|)
=\frac{\sqrt{1-|a|^2}}{\pi(1-v_1^2)
\sqrt{|a|^2-v_1^2}}
{\bf 1}(|v_1| < |a|), \\
\label{eqn:M1}
&& {\cal M}^{(1)}(A^{(1)}, \q)
=1-\left(|q_1|^2-|q_2|^2
+\frac{q_1 \overline{q}_2 a \overline{b}
+\overline{q}_1 q_2 \overline{a} b}{|a|^2} \right).
\end{eqnarray}
Here ${\bf 1}(\omega)$ denotes the indicator function
of a condition $\omega$;
${\bf 1}(\omega)=1$ if $\omega$ is satisfied
and ${\bf 1}(\omega)=0$ otherwise.
Figure \ref{fig:fig_SatoKatori_1}
shows the density function $\mu^{(1)}(v_1; |a|)$, 
when $a=1/\sqrt{2}$.
The function $\mu^{(1)}$ is now called 
the {\it Konno density function} \cite{BBCKPX08}.

\begin{figure}[htpb]
\begin{center}
\includegraphics[width=0.8\linewidth]{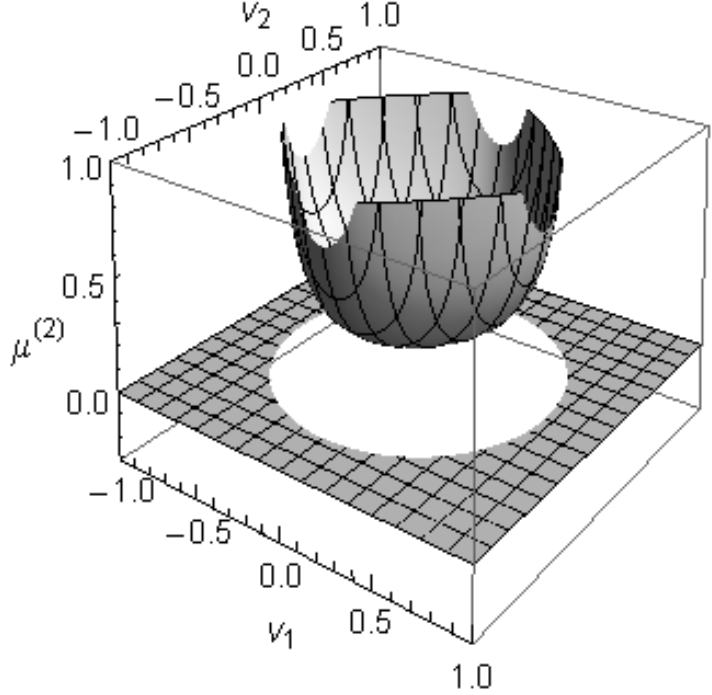}
\caption{
The two-dimensional density function
$\mu^{(2)}(v_1, v_2; p)$
of limit distribution of pseudovelocity
for the SQW$^{(2)}$, when $p=1/2$.}
\label{fig:fig_SatoKatori_2}
\end{center}
\end{figure}

Recently the weak convergence theorem of the
two-component pseudovelocity
\begin{equation}
\V(t) = (V_1(t), V_2(t))
= \left( \frac{X_1(t)}{t}, \frac{X_2(t)}{t} \right)
\label{eqn:vel2}
\end{equation}
for the SQW$^{(2)}$, where $\X(t)=(X_1(t), X_2(t))$
denotes the position of the walker, 
was also established \cite{WKKK08},
where the quantum die is parameterized by
$p \in (0, 1)$ as
\begin{equation}
A^{(2)}=
\left(\begin{array}{cccc}
-p & 1-p & \sqrt{p(1-p)} & \sqrt{p(1-p)}\\
1-p & -p & \sqrt{p(1-p)} & \sqrt{p(1-p)}\\
\sqrt{p(1-p)} & \sqrt{p(1-p)} & -(1-p) & p\\
\sqrt{p(1-p)} & \sqrt{p(1-p)} & p & -(1-p)
\end{array} \right) \in {\rm U}(4).
\label{eqn:A2}
\end{equation}
For any $\alpha_1, \alpha_2 \in \N_0$,
\begin{equation}
\lim_{t \to \infty}
\Big\langle V_1(t)^{\alpha_1} V_2(t)^{\alpha_2} \Big\rangle
=\int_{-\infty}^{\infty} d v_1 \int_{-\infty}^{\infty} d v_2
\, v_1^{\alpha_1} v_2^{\alpha_2}
\nu^{(2)}(v_1, v_2; A^{(2)}, \q)
\label{eqn:limit2d}
\end{equation}
where
\begin{equation}
\nu^{(2)}(v_1, v_2; A^{(2)}, \q)
=\mu^{(2)}(v_1, v_2; p) {\cal M}^{(2)}(v_1, v_2; A^{(2)}, \q)
\label{eqn:nu2}
\end{equation}
with
\begin{eqnarray}
\mu^{(2)}(v_1, v_2; p)
&=& \frac{2}{\pi^{2} 
(v_1+v_2+1)(v_1-v_2+1)(v_1+v_2-1)(v_1-v_2-1)} 
\nonumber\\
&& \qquad \times
 {\bf 1}(v_1^{2}/p+v_2^{2}/(1-p) < 1).
\label{eqn:mu2}
\end{eqnarray}
The explicit expression of ${\cal M}^{(2)}(v_1,v_2; A^{(2)}, \q)$
as a function of $\v=(v_1,v_2)$, 
the parameter $p$ of the quantum die
$A^{(2)}$, and the four-component initial qubit
$\q=(q_1, \dots, q_4)$ is given in Sec.III.B in \cite{WKKK08}.
Figure \ref{fig:fig_SatoKatori_2}
shows the $p=1/2$ case of the density function
$\mu^{(2)}$, which will be a two-dimensional extension
of the Konno density function (\ref{eqn:mu1}).
The common feature of $\mu^{(1)}$ and $\mu^{(2)}$
is that they are inverted-bell shaped on bounded supports,
an interval $(-a, a)$ and 
an elliptic region $\{(v_1, v_2) : v_1^2/p+v_2^2/(1-p) < 1\}$,
respectively, as shown in Figs.\ref{fig:fig_SatoKatori_1}
and \ref{fig:fig_SatoKatori_2}.
It is in a big contrast with
the Gaussian distributions in $d=1$ and $d=2$, 
which are bell shaped with unbounded supports
$\R$ and $\R^2$ describing
the diffusion scaling limits of the SRW models,
the one- and two-dimensional Brownian motions.

The purpose of the present paper is to clarify the physical
meaning of the pseudovelocities of quantum walkers,
which can have limit distributions in the weak sense
even in the quantum systems, 
and the origin of the inverted-bell shape of
the limit distribution with bounded supports.
Our argument is based on the
resemblance between the quantum walk models
and time-evolutions of multi-component wave functions
in the {\it relativistic quantum mechanics}
\cite{KFK05,Str06,BES07,Str07}.
Since $V^{(d)}(\k)$ given by (\ref{eqn:V1}) is 
a unitary matrix, we can assign a Hermitian matrix
${\cal H}^{(d)}(\k)$ such that
\begin{equation}
V^{(d)}(\k)=e^{-\im {\cal H}^{(d)}(\k)/\hbar},
\label{eqn:Hamil1}
\end{equation}
where $\hbar$ is a constant, the Planck constant
divided by $2 \pi$,
and then the time evolution (\ref{eqn:Psik})
of the SQW can be regarded as a solution of the equation
\begin{equation}
\im \hbar \frac{\partial}{\partial t}
\widehat{\Psi}(\k,t)={\cal H}^{(d)}(\k) 
\widehat{\Psi}(\k,t).
\label{eqn:Sch}
\end{equation}
Here time $t$ is now thought to be a continuous
variable in $\R$.

In an earlier paper \cite{KFK05}, a relation
was reported 
between the SQW$^{(1)}$ model and the Weyl equation,
which is given by (\ref{eqn:Sch}) for the two
component wave function,
$\widehat{\Psi}(\k,t)={^{\rm t}}(\psi_1(\k,t) \,
\psi_2(\k,t))$ with the Hamiltonian
${\cal H}^{(2)}(\k)=\vsigma \cdot \k$, where
$\vsigma=(\sigma_1, \sigma_2, \sigma_3)$ are the
Pauli matrices,
\begin{equation}
\sigma_1=\left( \begin{array}{cc} 
0 & 1 \cr 1 & 0 \end{array} \right), \quad
\sigma_2=\left( \begin{array}{cc} 
0 & -\im \cr \im & 0 \end{array} \right), \quad
\sigma_3=\left( \begin{array}{cc} 
1 & 0 \cr 0 & -1 \end{array} \right).
\label{eqn:Pauli1}
\end{equation}
Similarly, we expect a relation between the
SQW$^{(2)}$ and the Dirac equation,
which is given in the form of (\ref{eqn:Sch})
for the four-component wave function
$\widehat{\Psi}(\k,t)={^{\rm t}}(\psi_1(\k,t) \, \dots,
\psi_4(\k,t))$.
Note that the Weyl equation and the Dirac equation
are the time-evolution equations
for massless and massive relativistic particles
in quantum mechanics,
both in the three-dimensional continuous space
$\R^3$.
So we have to note that
the dimensionality $d$ of the SQW$^{(d)}$
and that of the corresponding relativistic quantum mechanics
is different from each other in general \cite{KFK05}.
Moreover, we have to introduce a proper
{\it ultraviolet cutoff} in the $\k$-space
for the quantum mechanics, in order to
establish the connection with the SQW$^{(d)}$ 
as explained below in detail.

In Sec.II first we introduce the Dirac equation
for a free particle in a usual way \cite{Dir58,BD64,IZ80}.
Then we show that it can provide
a model describing motion of 
a quantum walker having a four-component
qubit by identifying the time-evolution matrix (\ref{eqn:V1})
with the Hamiltonian matrix for the Dirac equation explicitly.
Then we study the long-time behavior of the moments
of pseudovelocity for a free Dirac particle
starting from the origin.
Our initial state will be considered as
an ideal limit of the highly localized state
studied by Bracken, Flohr, and Melloy \cite{BFM05}.
In this situation  we
clarify the fact that the pseudovelocity
is exactly equal to the {\it relativistic velocity}.
Since the Dirac particle as well as our quantum walker
should be {\it a tardyon, not a tachyon},
its velocity is less than the speed of light $c$,
\begin{equation}
v=\sqrt{v_1^2+v_2^2+v_3^2} < c.
\label{eqn:light}
\end{equation}
This is the physical origin of the finiteness
of supports in the limit distributions of $\V(t)$ universally
found in quantum walk models.
In order to ensure the convergence of integrals
giving joint moments of $V_j(t), 1 \leq j \leq 3,$ in $t \to \infty$,
we have to introduce an ultraviolet
cutoff in the theory, which will correspond to
the fact that quantum walk models are defined
on discrete spaces such as, lattices and trees.
The introduction of an ultraviolet cutoff here
will be equivalent with replacement of the 
delta-function-like initial state
by a wave packet with a finite size as studied by
Strauch \cite{Str07}
for the one-dimensional systems (see Sec.III C).
In Sec.III, we modify the Dirac equations
by reducing the number of components of momentum
to describing the quantum walk models in 
the dimensions lower than $d=3$.
We list up the limit distributions
of pseudovelocities of quantum walkers
corresponding to these free Dirac particles
in the dimensions $d=1$ and 2 with a proper
ultraviolet cutoff.
Then we compare the results
with the limit distributions of the SQW$^{(d)}$
with $d=1$ and 2 obtained 
in the previous papers \cite{Kon02,Kon05,KFK05,WKKK08}.
Concluding remarks are given in Sec.IV
associated with Appendix A.

\section{LIMIT DISTRIBUTION OF DIRAC EQUATION}
\subsection{Dirac equation as a quantum walk model}

The Dirac equation for a free particle 
with the rest mass $m$ in the three-dimensional 
real space $\R^3$ is 
given by \cite{Dir58,BD64,IZ80}
\begin{equation}
\im \hbar \frac{\partial}{\partial t}\Psi(\x,t) 
=\widehat{\cal H}{\Psi}(\x,t)
\label{eqn:Dirac1}
\end{equation}
for the four-component wave function $\Psi(\x,t)$
with the Hamiltonian operator
\begin{equation}
\widehat{\cal H}=
\gamma_4 \left(
\sum_{k=1}^{3} c \hbar 
{\gamma_k}\frac{\partial}{\partial x_k}+mc^2 
\right),
\label{eqn:DiracH}
\end{equation}
where ${\gamma_\nu}, \nu=1,2,3,4$, are the $4\times4$ gamma matrices
satisfying the algebra
\begin{equation}
{\gamma_\mu}{\gamma_\nu}+{\gamma_\nu}{\gamma_\mu}=2\delta_{\mu \nu}
I_4,
\quad \mu, \nu=1,2,3,4
\label{eqn:gamma1}
\end{equation}
with the $4 \times 4$ unit matrix $I_4$.
In the present paper, we fix the matrix representations
as follows,
\begin{equation}
\gamma_{k}=\left( \begin{array}{cc}
\0 & - \im \sigma_k \cr \im \sigma_k & \0 
\end{array} \right), \quad k=1,2,3,
\quad
\gamma_4= \left( \begin{array}{cc}
I & \0 \cr \0 & -I
\end{array} \right),
\label{eqn:gamma3}
\end{equation}
where $\sigma_k, k=1,2,3$, are the Pauli matrices
given by (\ref{eqn:Pauli1}) and
$I$ and $\0$ are the $2 \times 2$ unit matrix
and the $2 \times 2$ zero matrix, respectively.
From now on we consider the
momentum $\p=(p_1, p_2, p_3)$ of
the particle, instead of the
wave number vector $\k=(k_1, k_2, k_3)$
of the wave function, where
the de Broglie relation $\p=\hbar \k$
is established.
The momentum $\p$ provides a good quantum number and
the solution for a given momentum 
$\p$ is given by a plane wave,
\begin{equation}
\Psi_{\p}(\x,t)=
e^{\im \p \cdot \x/\hbar
- \im E(p)t/\hbar} u(\p), 
\label{eqn:planewave}
\end{equation}
where 
\begin{equation}
p=|\p|=\sqrt{p_1^2+p_2^2+p_3^2}, \quad
E(p)=\sqrt{(pc)^2 +(mc^2)^2}, 
\label{eqn:Ep}
\end{equation}
and $u(p)$ is a four-component vector which satisfies
the eigenvalue equation
\begin{equation}
E(p) u(\p)={\cal H}(\p) u(\p)
\label{eqn:Dirac2}
\end{equation}
with the Hamiltonian matrix
\begin{equation}
{\cal H}(\p)=
{\gamma_4} \left( \im \sum_{k=1}^{3} c{\gamma_k}p_k+mc^2 \right).
\label{eqn:DiracH2}
\end{equation}
Then, in the momentum space, the wave function is given as
\begin{equation}
\widehat{\Psi}(\p,t)= e^{-\im {\cal H}(\p) t/\hbar} u(\p).
\label{eqn:up1}
\end{equation}

The Hamiltonian matrix given by 
(\ref{eqn:DiracH2}) can be diagonalized 
by the Foldy-Wouthuysen-Tani transformation 
\cite{FW50,Tani51} as follows: 
\begin{equation}
 {\cal H}(\p)=U(\p)^{-1} E(p) \ \gamma_4 U(\p)
\end{equation}
with (\ref{eqn:Ep}) and
\begin{eqnarray}
U(\p) &=& \frac{1}{\sqrt {2E(p)}}
\left(\sqrt{E(p)+mc^2} \, I_4 +\im \sum_{k=1}^{3}
\frac{c \gamma_k p_k}{\sqrt{E(p)+mc^2}} \right)
\nonumber\\
&=& \frac{1}{\sqrt{2E(p)}}\left(
           \begin{array}{cccc}
\sqrt{E(p)+mc^2} & 0 & \frac{c p_3}{\sqrt{E(p)+mc^2}}
& \frac{c(p_1-\im p_2)}{\sqrt{E(p)+mc^2}}  \\
0 & \sqrt{E(p)+mc^2} & \frac{c(p_1+ \im p_2)}{\sqrt{E(p)+mc^2}} 
& \frac{-c p_3}{\sqrt{E(p)+mc^2}} \\
\frac{-cp_3}{\sqrt{E(p)+mc^2}} & \frac{-c(p_1- \im p_2)}{\sqrt{E(p)+mc^2}} 
& \sqrt{E(p)+mc^2} & 0  \\
\frac{-c(p_1+ \im p_2)}{\sqrt{E(p)+mc^2}} 
& \frac{c p_3}{\sqrt{E(p)+mc^2}} & 0 & \sqrt{E(p)+mc^2} 
\end{array}
\right).
\label{eqn:U1}
\end{eqnarray}
Then we can see that
\begin{eqnarray}
e^{-\im {\cal H}(\p)t/\hbar}               
&=& \sum_{n=0}^{\infty} \frac{1}{n!} 
\left(-\im \frac{{\cal H}(\p)}{\hbar}t\right)^n
\nonumber\\
&=& \sum_{n=0}^{\infty} \frac{1}{n!}
\left(\frac{- \im t}{\hbar}\right)^n 
(U(\p)^{-1} E(p) \ \gamma_4 U(\p))^n
\nonumber\\
&=& U(\p)^{-1} \left[\sum_{n=0}^{\infty} 
\frac{1}{n!}\left(\frac{- \im t}{\hbar}\right)^n 
(E(p) \ \gamma_4 )^n \right] U(\p)
\nonumber\\
&=& \left[U(\p)^{-1} 
{\rm diag} \Big[
e^{-\im E(p)/\hbar}, e^{-\im E(p)/\hbar}, e^{\im E(p)/\hbar},
e^{\im E(p)/\hbar} \Big] U(\p) \right]^t.
\label{eqn:calc1}
\end{eqnarray}
It implies that the Dirac equation provides
a quantum walk model,
in which the walker is driven by
the $4 \times 4$ unitary matrix
\begin{equation}
V(\p)=U(\p)^{-1} 
{\rm diag} \Big[
e^{-\im E(p)/\hbar}, e^{-\im E(p)/\hbar}, e^{\im E(p)/\hbar},
e^{\im E(p)/\hbar} \Big] U(\p),
\label{eqn:Vp1}
\end{equation}
where $U(\p)$ is given by (\ref{eqn:U1}).

\subsection{Limit distribution of
relativistic velocity}

Let $\X(t)=(X_1(t), X_2(t), X_3(t))$ be the
position of a free Dirac particle at time $t$
starting from the origin at time $t=0$.
The joint moments of $X_j(t), 1 \leq j \leq 3$ are given by 
\begin{equation}
\left\langle \prod_{j=1}^{3} X_j(t)^{\alpha_j} \right\rangle
= \prod_{j=1}^{3} \int_{-\infty}^{\infty}\frac{dp_j}{2\pi\hbar} \,
\widehat{\Psi}^{\dagger}(\p,t)
\prod_{j=1}^{3}
\left(\im \hbar\frac{\partial}{\partial p_j}\right)^{\alpha_j} 
\widehat{\Psi}(\p,t).
\label{eqn:momentD1}
\end{equation}
Let $\w_j(\p)$ be the $j$-th row of the matrix $U(\p)$
given by (\ref{eqn:U1}), $1 \leq j \leq 4$.
We assume the initial state as
\begin{equation}
\widehat{\Psi}(\p,0)= \widehat{\Psi}_{0} \equiv
{^{\rm t}} ( \begin{array}{cccc}
q_1 & q_2 & q_3 & q_4 
\end{array}),
\label{eqn:initial2}
\end{equation}
where $q_j \in \C, 1 \leq j \leq 4$, are
independent of $\p$ and $\sum_{j=1}^{4} |q_j|^2=1$.
Then
\begin{eqnarray}
\widehat{\Psi}(\p,t) 
&=& \left[
U(\p)^{-1}
{\rm diag} \Big[
e^{-\im E(p)/\hbar}, e^{-\im E(p)/\hbar}, e^{\im E(p)/\hbar},
e^{\im E(p)/\hbar} \Big] U(\p) \right]^{t} \widehat{\Psi}_0    
\nonumber \\
&=&e^{-\im E(p)t/\hbar} 
\{ \w_1(\p) C_1(\p)+ \w_2(\p)C_2(\p) \}  
\nonumber\\
&&+e^{\im E(p)t/\hbar} 
\{ \w_3(\p)C_3(\p)+ \w_4(\p)C_4(\p) \},
\label{eqn:Psit1}
\end{eqnarray}
where 
$C_j(\p)\equiv [\w_j(\p)]^\dag \widehat\Psi_0, 1 \leq j \leq 4$. 

For $\alpha_j \in \N \equiv \{1,2, \cdots\}, 
1 \leq j \leq 3$, we see \cite{GJS04}
\begin{eqnarray}
&& \prod_{j=1}^{3}
 \left(\im 
 \hbar\frac{\partial}{\partial p_j} \right)^{\alpha_j} 
 \widehat{\Psi}(\p,t) \nonumber\\
&=& \prod_{j=1}^{3}
\left(\frac{\partial E(p)}{\partial p_j} \right)^{\alpha_j} 
e^{-\im E(p)t/\hbar}
\Big[\w_1(\p) C_1(\p)+\w_2(\p) C_2(\p) \Big] 
    {t}^{\sum_{j=1}^{3} \alpha_j}
\nonumber\\
&+& 
\prod_{j=1}^{3}
\left(-\frac{\partial E(p)}{\partial p_j}\right)^{\alpha_j} 
e^{\im E(p)t/\hbar}
\Big[\w_3(\p) C_3(\p)+\w_4(\p) C_4(\p) \Big] 
    {t}^{\sum_{j=1}^{3} \alpha_j}
 +{\cal O} \left(t^{\sum_{j=1}^{3} \alpha_j-1} \right). 
\label{eqn:exp1}
\end{eqnarray}
Since $U(\p)$ is unitary, its row vectors 
$\{\w_j(\p)\}_{j=1}^{4}$ 
make a set of orthonormal vectors, 
\begin{equation}
\w_j^{\dag}(\p) \w_{k}(\p)=\delta_{j k},
\quad 1 \leq j, k \leq 4.
\label{eqn:orto}
\end{equation}
Then we have
\begin{eqnarray}
&& \widehat{\Psi}^{\dag}(\p,t)
\prod_{j=1}^{3} 
\left(\im \hbar\frac{\partial}{\partial p_j}\right)^{\alpha_j} 
\widehat{\Psi}(\p,t) =
\prod_{j=1}^{3} 
\left(\frac{\partial E(p)}{\partial p_j}\right)^{\alpha_j} 
\nonumber\\
&& \qquad \times \left[ | C_1(\p) |^2+ |C_2(\p)|^2  
+(-1)^{\sum_{j=1}^{3} \alpha_j}
\{| C_3(\p) |^2+ |C_4(\p)|^2 \} \right]
{t}^{\sum_{j=1}^{3} \alpha_j}
\nonumber\\
&& \qquad \qquad 
+{\cal O}\left(t^{\sum_{j=1}^{3} \alpha_j-1} \right). 
\label{eqn:exp2}
\end{eqnarray}
The pseudovelocity is defined as 
$$
\V(t)=\left(\frac{X_1(t)}{t},\frac{X_2(t)}{t},
\frac{X_3(t)}{t} \right),
$$
and we obtain the long-time limit of the joint moments; 
for $\alpha_j \in \N_0, 1 \leq j \leq 3$
\begin{eqnarray}
& &\lim _{t \to \infty} 
\left\langle \prod_{j=1}^{3} V_j(t)^{\alpha_j} 
\right\rangle \nonumber \\
&=& \prod_{j=1}^{3} \int_{-\infty}^{\infty}\frac{dp_j}{2\pi\hbar}  
 \left[\{ | C_1(\p) |^2+ |C_2(\p)|^2 \} 
 +(-1)^{\sum_{j=1}^{3} \alpha_j}
 \{| C_3(\p) |^2+ |C_4(\p)|^2\} \right]  \nonumber\\
& & \qquad 
\times \prod_{j=1}^{3} 
\left(\frac{\partial E(p)}{\partial p_j}\right)^{\alpha_j}.
\label{eqn:exp3}
\end{eqnarray}
Since $E(p)$ is the relativistic energy of
the particle given by (\ref{eqn:Ep}), 
\begin{equation}
\frac{\partial E(p)}{\partial p_j}
=c^2 \frac{p_j}{E(p)}, \quad j=1,2,3.
\label{eqn:ptov1}
\end{equation}
We change the variables of integrals in (\ref{eqn:exp3}) from
$p_j$'s to $v_j$'s by the transformation
\begin{equation}
v_j=c^2 \frac{p_j}{E(p)}, \quad j=1,2,3.
\label{eqn:ptov}
\end{equation}
We can show that this map 
$\p \in \R^3 \mapsto \v=(v_1,v_2,v_3)$
is one-to-one and the image is an interior of a circle
with radius $c$,
\begin{equation}
v_x^2+v_y^2+v_z^2 <c^2 . 
\label{eqn:image}
\end{equation}
Moreover, we find that the following relations are 
equivalent with (\ref{eqn:ptov}), 
\begin{equation}
 p_j= \frac{mv_j}{\sqrt{1-v^2/c^2}},
 \quad j=1,2,3,
 \label{eqn:ptov2}
\end{equation}
where $v=|\v|$.
That is, $\v$ is the relativistic velocity
of the particle.
From this expression (\ref{eqn:ptov2}) 
we can readily calculate
the Jacobian associated with the inverse map
$\v \mapsto \p$,
$$
J \equiv  \det \left[
\frac{\partial v_j}{\partial p_k} \right]_{1 \leq j,k \leq 3}
 =c^{10}\frac{m^2}{E(p)^5}
 =\frac{1}{m^3} \left( 
 {1-\frac{v^2}{c^2}} \right)^{5/2}. 
$$
Associated with the change of variables,
$C_j(\p)$'s are replaced by 
$\widehat{C}_j(\v)$'s and the integrals 
in (\ref{eqn:exp3}) are rewritten as 
\begin{eqnarray}
& &\lim _{t \to \infty} 
\left\langle \prod_{j=1}^{3} V_j(t)^{\alpha_j} 
\right\rangle \nonumber \\
&=& \prod_{j=1}^{3} \int_{-\infty}^{\infty}
\frac{dv_j}{2\pi\hbar}  
\frac{1}{J}
 \left[\{ | \widehat{C}_1(\v) |^2+ |\widehat{C}_2(\v)|^2 \} 
 +(-1)^{\sum_{j=1}^{3} \alpha_j}
 \{| \widehat{C}_3(\v) |^2+ |\widehat{C}_4(\v)|^2\} \right]  \nonumber\\
& & \qquad \qquad 
\times \prod_{j=1}^{3} v_j^{\alpha_j}  
{\bf 1}(v^2_1+v^2_2+v^3_3 <c^2) \nonumber   \\
&=& \prod_{j=1}^{3} \int_{-\infty}^{\infty} \frac{dv_j}{c} 
    \prod_{j=1}^{3} v_j^{\alpha_j} 
    \mu_{\rm Dirac}^{(3)}(v){\cal M}_{\rm Dirac}^{(3)}(\v;\q)
\label{eqn:limitDirac1}
\end{eqnarray} 
for $\alpha_j \in \N_0, 1 \leq j \leq 3$.
Here 
\begin{equation}
\mu_{\rm Dirac}^{(3)}(v)
     =\left(\frac{m c}{2\pi\hbar}\right)^3
     \frac{1}{(1-v^2/c^2)^{5/2}}
     {\bf 1}(v < c),
\label{eqn:muDirac}
\end{equation} 
and
\begin{equation}
{\cal M}_{\rm Dirac}^{(3)}(\v; \q)=
1+\sum_{j=1}^{3}{\cal M}_{\rm Dirac}^{^{(3,j)}}(\q) v_j
\label{eqn:MDirac}
\end{equation}
with
\begin{eqnarray}
 && {\cal M}_{\rm Dirac}^{(3,1)}(\q) 
 =2 {\rm Re}(q_1\bar{q_4}+q_2\bar{q_3}) \quad
 {\cal M}_{\rm Dirac}^{(3,2)}(\q)
 =2 {\rm Im}(\bar{q_1}q_4-\bar{q_2} q_3),
 \nonumber\\
 &&
 {\cal M}_{\rm Dirac}^{(3,3)}(\q)=2 {\rm Re}(q_1\bar{q_3}-q_2\bar{q_4}),
\label{eqn:MDirac2}
\end{eqnarray}
where ${\rm Re}(z)$ and ${\rm Im}(z)$
denote the real part and the imaginary part of $z \in \C$,
respectively. 

\subsection{Ultraviolet cutoff}

The integrals
\begin{equation}
 \prod_{j=1}^{3}
\int_{-\infty}^{\infty} \frac{dv_j}{c}
\prod_{j=1}^{3} v_j^{\alpha_j}
\mu_{\rm Dirac}^{(3)}(v) {\cal M}^{(3)}_{\rm Dirac}(\v;\q),
\quad (\alpha_1, \alpha_2, \alpha_3) \in \N_0^3
\label{eqn:diverge}
\end{equation}
generally do not converge. 
In order to obtain finite values of physical quantities
the momentum space should be
restricted, 
and we set an ultraviolet cutoff by
introducing a cutoff parameter $\lambda >0$ as
\begin{equation}
p=|\p| < \lambda.
\label{eqn:cutoff}
\end{equation}
The range of $\v=(v_1,v_2,v_3)$ is then
\begin{equation}
v_1^2+v_2^2+v_3^2 < \frac{\lambda^2c^2}{(mc)^2+\lambda^2}.
\label{eqn:cutoff2}
\end{equation}
We can calculate the normalization constant for
finite $\lambda$ as
\begin{equation}
\prod_{j=1}^{3} \int_{-\infty}^{\infty} \frac{dv_j}{c}
\left(\frac{m c}{2\pi\hbar}\right)^3
\frac{1}{(1-v^2/c^2)^{5/2}}
{\bf 1} 
\left(v < \sqrt{\frac{\lambda^2c^2}{(mc)^2+\lambda^2}} \right)
=\frac{ \lambda^3}{6 \pi^2\hbar^3}.
\label{eqn:cutoff3}
\end{equation}
Finally the distribution function is determined as
\begin{equation}
\mu_{\rm Dirac}^{(3)}(v; \lambda)=
\frac{3}{4 \pi} \left(\frac{m c}{\lambda}\right)^3 
\frac{1}{(1-v^2/c^2)^{5/2}}
{\bf 1} \left( v < \sqrt{
\frac{\lambda^2c^2}{(mc)^2+\lambda^2}} \right).
\label{eqn:muDirac2}
\end{equation}

\begin{figure}[htpb]
\begin{center}
\includegraphics[width=0.4\linewidth]{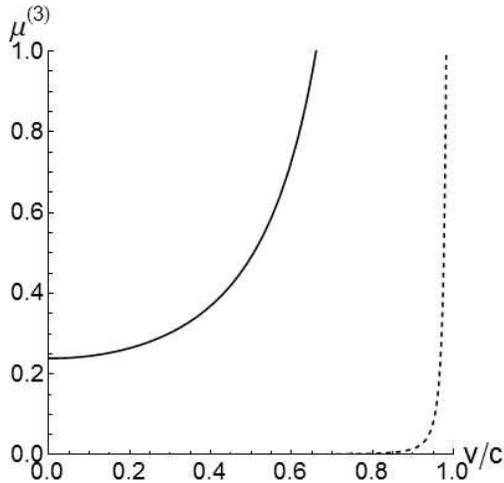}
\caption{
The dependence of $\mu_{\rm Dirac}^{(3)}$
on the magnitude $v$ of the pseudovelocity.
The solid line for the cutoff
parameter $\lambda/(mc)=1$,
and the broken line for $\lambda/(mc)=10$,
respectively.}
\label{fig:fig_SatoKatori_3}
\end{center}
\end{figure}

We assume that $m$ is the rest mass
of an electron.
Figures \ref{fig:fig_SatoKatori_3} shows 
$\mu^{(3)}_{\rm Dirac}(v;\lambda)$
given by (\ref{eqn:muDirac2})
as functions of $v=\sqrt{v_1^2+v_2^2+v_3^2} >0$ 
for the cutoff parameter
$\lambda/(mc)=1$ (by the solid line) 
and 10 (by the broken line), respectively.
They should be compared with Fig.1 in \cite{BFM05}.
If $\lambda/(mc) \to \infty$,
that is,  the ultraviolet cutoff of the energy
$\overline{E}=\lambda c$ becomes much greater than
the rest mass energy $mc^2$ of the particle,
$\mu_{\rm Dirac}^{(3)}(v; \lambda) \to \delta(v-c)$;
the velocity of spreading of the 
Dirac particle outwards from the origin 
tends to be close to the speed of light $c$
\cite{BFM05,Str06,BES07}.
Since quantum walk models are defined on lattices,
an ultraviolet cutoff is naturally introduced
so that $\lambda/(mc)=\overline{E}/(mc^2)$ 
remains to be 
small, and thus the inverted-bell shaped distributions
of pseudovelocities are universally observed.

The reason that the integrals for the moments (\ref{eqn:diverge})
do not converge is due to our choice of initial state
such that a Dirac particle is put at the origin at time $t=0$.
It will be equivalent with setting the initial wave function
be a delta function at the origin, which is not
square-integrable.
From this point of view, the introduction of ultraviolet
cutoff in the present paper can be regarded as
a modification of the initial state.
In the paper \cite{Str07}, Strauch introduced 
a {\it localization parameter} $a$, which controls 
the effective size of the initial wave packet describing
a Dirac particle and 
a quantum walker in the models.
Dependence of the spatial distributions
of wave packets in $t > 0$ on the parameter $a$ and 
``initial qubit" was fully studied for the
one-dimensional Dirac equation and the quantum walk models.
We hope that the present paper will show the
importance of his study to understand the relationship
between the quantum walk models and the relativistic
quantum mechanics also in the higher spatial-dimensions.

\section{LOWER DIMENSIONAL SYSTEMS}
\subsection{Two-dimensional system}
We consider the Hamiltonian matrix
\begin{equation}
 {\cal H}(\p)=U(\p)^{-1}
 \sqrt{(pc)^2+(mc^2)^2} \ \gamma_4 U(\p)
\end{equation}
with the two-component momentum
$\p=(p_1,p_2)$, where 
$p=|\p|=\sqrt{p_1^2+p_2^2}$ and
\begin{eqnarray*}
& &U(\p) 
=\frac{1}{\sqrt {2E(p)}} \left(\sqrt{E(p)+mc^2} \, I_4
+\im \sum_{k=1}^{2} \frac{c \gamma_k p_k}{\sqrt{E(p)+mc^2}} \right) \\
& & =\frac{1}{\sqrt{2E(p)}}\left(
   \begin{array}{cccc}
   \sqrt{E(p)+mc^2} & 0 & 0 & \frac{c(p_1- \im p_2)}{\sqrt{E(p)+mc^2}}  \\
   0 & \sqrt{E(p)+mc^2} & \frac{c(p_1+\im p_2)}{\sqrt{E(p)+mc^2}} & 0 \\
   0 & \frac{-c(p_1-\im p_2)}{\sqrt{E(p)+mc^2}} & \sqrt{E(p)+mc^2} & 0  \\
   \frac{-c(p_1+ \im p_2)}{\sqrt{E(p)+mc^2}} & 0 & 0 & \sqrt{E(p)+mc^2} 
\end{array}
\right)
\end{eqnarray*}
with 
$E(p)=\sqrt{(pc)^2+(mc^2)^2}$.
Following the similar procedure to that given in Sec.II, 
we will obtain the
following result under the ultraviolet cutoff (\ref{eqn:cutoff})
with a parameter $\lambda$; for $\alpha_j \in \N_0, j=1,2$,
\begin{equation}
\lim_{t \to \infty} 
\left\langle \prod_{j=1}^{2} V_j(t)^{\alpha_j} \right\rangle_{\lambda}
= \prod_{j=1}^{2} \int_{-\infty}^{\infty} 
\frac{dv_j}{c} \,
\prod_{j=1}^{2} v_j^{\alpha_j} 
\nu_{\rm Dirac}^{(2)}(\v; \lambda, \q)
\label{eqn:2dimDirac0}
\end{equation}
with $v=\sqrt{v_1^2+v_2^2}$ and
\begin{equation}
\nu_{\rm Dirac}^{(2)}(\v; \lambda, \q)=
\mu_{\rm Dirac}^{(2)}(v; \lambda) 
{\cal M}_{\rm Dirac}^{(2)}(\v; \q),
\label{eqn:2dimDirac1}
\end{equation}
where
\begin{eqnarray}
\label{eqn:2dimDirac2}
&& \mu_{\rm Dirac}^{(2)}(v; \lambda)
     =\frac{1}{\pi} \left(\frac{mc}{\lambda}\right)^2  
      \frac{1}{(1-v^2/c^2)^2}
      {\bf 1} \left(
      v < \sqrt{\frac{\lambda^2 c^2}{(mc)^2+ \lambda^2}}
      \right), \\
\label{eqn:2dimDirac3}
&& {\cal M}_{\rm Dirac}^{(2)}(\v; \q)
=1+\sum_{j=1}^{2} {\cal M}_{\rm Dirac}^{(2,j)}(\q) v_j
\end{eqnarray}
with
\begin{equation}
 {\cal M}_{\rm Dirac}^{(2,1)}(\q)
 =2 {\rm Re}(q_1\bar{q_4}+q_2\bar{q_3}), \quad
 {\cal M}_{\rm Dirac}^{(2,2)}(\q)
 =2 {\rm Im}(\bar{q_1}q_4-\bar{q_2} q_3). 
\label{eqn:2dimDirac4}
\end{equation}

\begin{figure}[htpb]
\begin{center}
\includegraphics[width=0.8\linewidth]{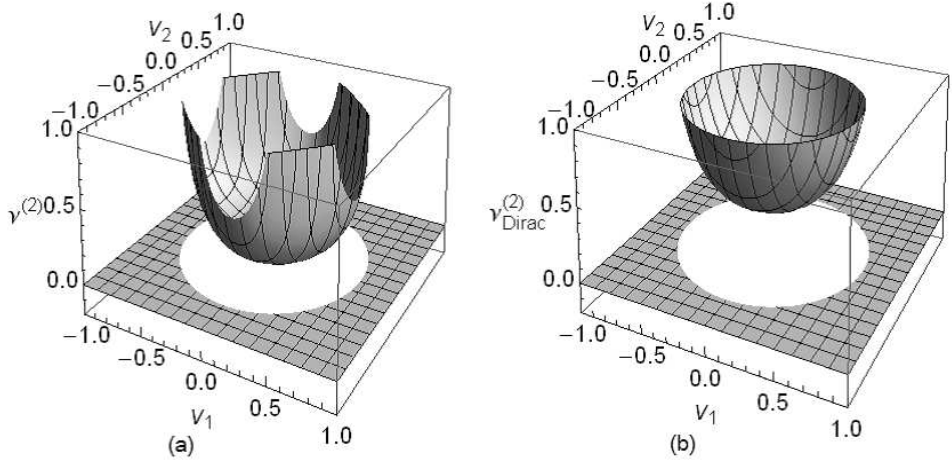}
\caption{
Comparison between 
(a) $\nu^{(2)}$ of the SQW$^{(2)}$ 
and (b) $\nu^{(2)}_{\rm Dirac}$.
The parameters are chosen as
(a) $p=1/2, \q=(1/2, -\im/2, -1/2, \im/2)$
and (b) $\lambda/(mc)=1,
\q=(1/\sqrt{2}, 1/\sqrt{2}, 0, 0)$, respectively.}
\label{fig:fig_SatoKatori_4}
\end{center}
\end{figure}

\begin{figure}[htpb]
\begin{center}
\includegraphics[width=0.8\linewidth]{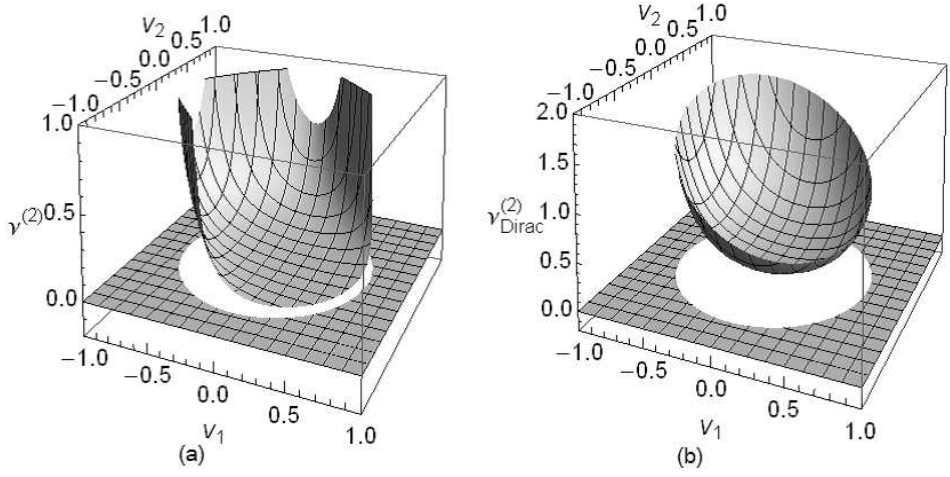}
\caption{
Comparison between 
(a) $\nu^{(2)}$ of the SQW$^{(2)}$
and (b) $\nu^{(2)}_{\rm Dirac}$.
The parameters are chosen as
(a) $p=1/2, \q=(1/2, \im/2, \im/2, -1/2)$
and (b) $\lambda/(mc)=1,
\q=(-(1+\im)/(2 \sqrt{2}), -(1+\im)/(2 \sqrt{2}), 
(1+\im)/(2 \sqrt{2}), (1-\im)/(2 \sqrt{2}))$, respectively.
}
\label{fig:fig_SatoKatori_5}
\end{center}
\end{figure}

Figure \ref{fig:fig_SatoKatori_4} shows 
the comparison between 
(a) $\nu^{(2)}$ of the SQW$^{(2)}$ given by (\ref{eqn:nu2})
(see \cite{WKKK08} for detail),
and (b) $\nu^{(2)}_{\rm Dirac}$ given by 
(\ref{eqn:2dimDirac1}).
Here we have chosen the parameters as
(a) $p=1/2, \q=(1/2, -\im/2, -1/2, \im/2)$
and (b) $\lambda/(mc)=1,
\q=(1/\sqrt{2}, 1/\sqrt{2}, 0, 0)$, respectively.
In this case, $\nu^{(2)}$ has the symmetry
$\nu^{(2)}(-v_1, -v_2)=\nu^{(2)}(v_1, v_2)$,
but it is not isotropic around the origin.
Since we have introduce a simple
isotropic cutoff in the momentum space (\ref{eqn:cutoff}),
$\nu^{(2)}_{\rm Dirac}$ is isotropic around the
origin. (See a remark in Sec.IV.)
Figure \ref{fig:fig_SatoKatori_5} shows 
the comparison between 
(a) $\nu^{(2)}$ with $p=1/2, \q=(1/2, \im/2, \im/2, -1/2)$
and (b) $\nu^{(2)}_{\rm Dirac}$
with $\lambda/(mc)=1,
\q=(-(1+\im)/(2\sqrt{2}), -(1+\im)/(2\sqrt{2}), 
(1+\im)/(2\sqrt{2}), (1-\im)/(2\sqrt{2}))$, respectively.
In this case, both distributions show
similar anisotropy around the origin.

\subsection{One-dimensional system}

By setting the momentum be a scalar $p_1$,
we can discuss the one-dimensional system
with a cutoff parameter $\lambda$.
The obtained limit distribution of the
one-component pseudovelocity is the following; 
for $\alpha_1 \in \N_0$
\begin{equation} 
\lim _{t \to \infty} 
\Big\langle V_1(t)^{\alpha_1} \Big\rangle_{\lambda}
= \int_{-\infty}^{\infty} \frac{dv_1}{c}   
{v_1}^{\alpha_1} \nu_{\rm Dirac}^{(1)}(v_1; \lambda, \q)
\label{eqn:1dimDirac0}
\end{equation}
with
\begin{equation}
\nu_{\rm Dirac}^{(1)}(v_1; \lambda, \q)=
\mu_{\rm Dirac}^{(1)}(v_1, \lambda) 
{\cal M}_{\rm Dirac}^{(1)} (v_1; \q),
\label{eqn:1dimDirac1}
\end{equation} 
where
\begin{eqnarray}
\label{eqn:1dimDirac2}
&& \mu_{\rm Dirac}^{(1)}(v_1, \lambda)
= \frac{mc}{2 \lambda} 
\frac{1}{(1-v_1^2/c^2)^{3/2}}
{\bf 1}\left(v_1 <
\sqrt{\frac{\lambda^2c^2}{(mc)^2+\lambda^2}} \right)
\\
\label{eqn:1dimDirac3}
&& 
{\cal M}_{\rm Dirac}^{(1)}(v_1;\q)
=1+2v_1 {\rm Re} (q_1\bar{q_4}+q_2\bar{q_3}).
\end{eqnarray}
The result (\ref{eqn:1dimDirac2}) can be compared with
Eq.(56) in \cite{BES07}.

\begin{figure}[htpb]
\begin{center}
\includegraphics[width=1.0\linewidth]{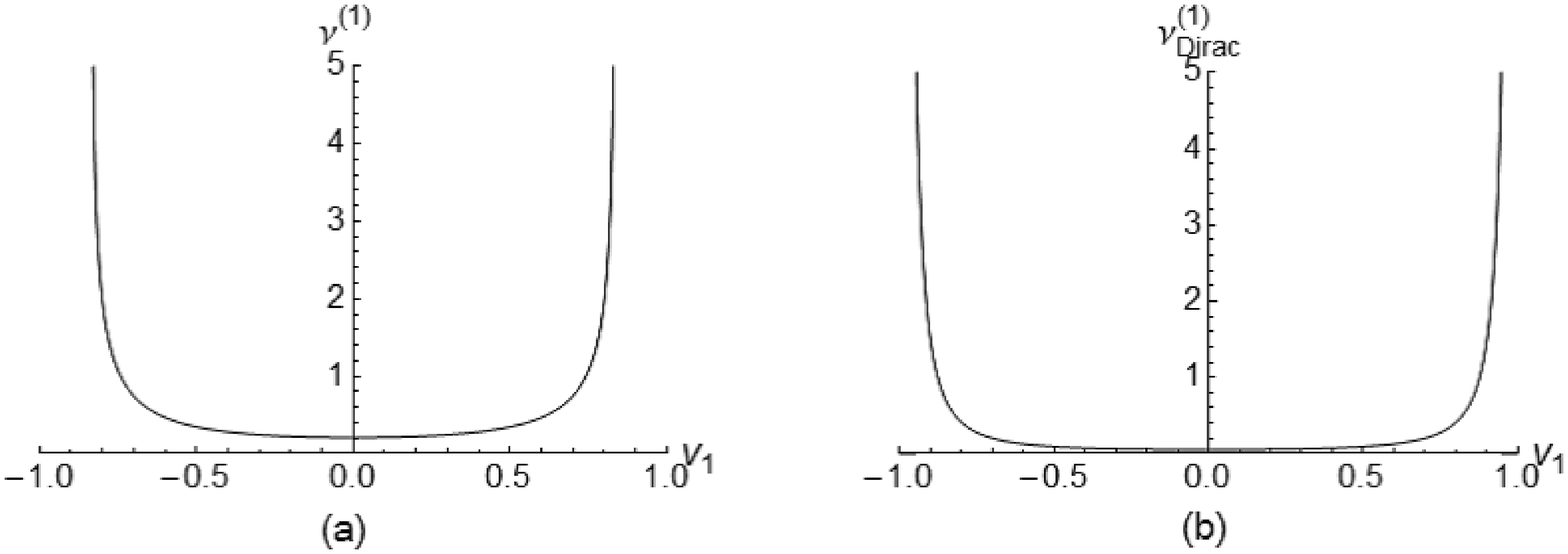}
\caption{
Comparison between 
(a) $\nu^{(1)}$ of the SQW$^{(1)}$
and (b) $\nu^{(1)}_{\rm Dirac}$.
The parameters are chosen as
(a) $a=\im \sqrt{7/10}, 
b=\im \sqrt{3/10}, \q=(1/\sqrt{2}, \im/\sqrt{2})$
and (b) $\lambda/(mc)=3,
\q=(1/\sqrt{2}, 1/\sqrt{2}, 0, 0)$, respectively.}
\label{fig:fig_SatoKatori_6}
\end{center}
\end{figure}

\begin{figure}[htpb]
\begin{center}
\includegraphics[width=1.0\linewidth]{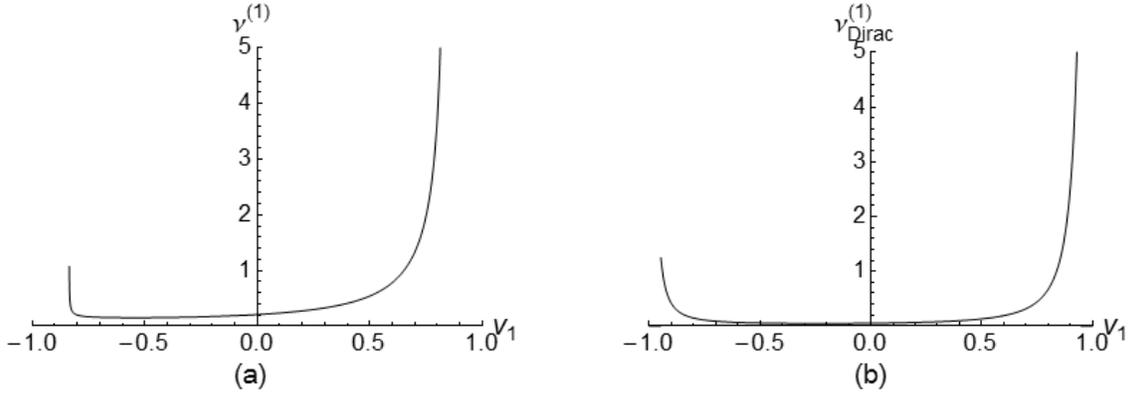}
\caption{
Comparison between 
(a) $\nu^{(1)}$ of the SQW$^{(1)}$
and (b) $\nu^{(1)}_{\rm Dirac}$.
The parameters are chosen as
(a) $a=\sqrt{7/10}, 
b=\im \sqrt{3/10}, \q=(1/\sqrt{5}, 2 \im/\sqrt{5})$
and (b) $\lambda/(mc)=3,
\q=(1/\sqrt{10},1/\sqrt{10},2/\sqrt{10},2/\sqrt{10})$, 
respectively.
}
\label{fig:fig_SatoKatori_7}
\end{center}
\end{figure}

Figure \ref{fig:fig_SatoKatori_6} shows 
the comparison between 
(a) $\nu^{(1)}$ of the SQW$^{(1)}$ given by (\ref{eqn:nu1})
and (b) $\nu^{(1)}_{\rm Dirac}$ given by 
(\ref{eqn:1dimDirac1}).
Here we have chosen the parameters as
(a) $a=\im \sqrt{7/10}, b=\im \sqrt{3/10}, 
\q=(1/\sqrt{2}, \im/\sqrt{2})$
and (b) $\lambda/(mc)=3,
\q=(1/\sqrt{2}, 1/\sqrt{2}, 0, 0)$, respectively.
Both distributions are symmetric.
Figure \ref{fig:fig_SatoKatori_7} shows 
the comparison between 
(a) $\nu^{(1)}$ with $a=\sqrt{7/10},
b=\im \sqrt{3/10}, 
\q=(1/\sqrt{5}, 2 \im/\sqrt{5})$
and (b) $\nu^{(1)}_{\rm Dirac}$
with $\lambda/(mc)=3,
\q=(1/\sqrt{10}, 1/\sqrt{10}, 2/\sqrt{10}, 2/\sqrt{10})$, respectively.
Both show similar asymmetry.

\section{CONCLUDING REMARKS}

In the present paper, we have shown that
the novel structures of the limit distributions of
pseudovelocities in the quantum walk models
can be explained by the theory of relativistic
quantum mechanics.
We have studied a very simple system,
a single free Dirac particle,
but we put two special setting when we solve the
Dirac equation, in order to clarify the connection
with the quantum walk models.
The first one is the initial condition such that the
particle is located at the origin at $t=0$, and 
the second one is the introduction of an
ultraviolet cutoff in the theory.

Importance of the study of a highly localized state
of a free Dirac particle, when we consider
the connection between the relativistic quantum mechanics
and the quantum walk models, was clearly
demonstrated by Bracken, Flohr, and Mellow \cite{BFM05},
by Strauch \cite{Str06}, 
and by Bracken, Ellinas, and Smyrnakis \cite{BES07}.
By observing the time evolution of the wave-packet-type
solutions of the Dirac equation, they found that
such a highly localized initial state of a free Dirac
particle leads to a rapid expansion of the distribution
of position of the particle outwards from the origin,
at speeds close to $c$.
Our initial condition will be regarded as 
an ideal limit of their setting, in which the width of
the wave-packet goes to zero.
If we study the long-time limit of
the pseudovelocity $\V(t)$ of the free Dirac particle,
this phenomenon gives a singular distribution
such that only on the surface of the sphere
with the radius $c$ we have the delta measure,
and nothing on any other points 
in the $\V$-space.
Since the quantum walk models studied so far
are defined on discrete spaces, such as lattices
and trees,
we have to introduce an ultraviolet cutoff
in the quantum mechanics.
Then the limit distributions of $\V(t)$
are moderated and we have obtained the inverted-bell shaped
distribution functions.
Introduction of an ultraviolet cutoff $\lambda < \infty$
in the momentum space (\ref{eqn:cutoff})
will be equivalent with the introduction of 
an effective size $a \propto \hbar/\lambda >0$
for the spatial initial state
of a quantum particle/walker.
For the one-dimensional models,
systematic study of the dependence
of solutions on the parameter $a$
was reported by Strauch \cite{Str07}
for the discrete- and the continuous-time
quantum walks and for the relativistic
and non-relativistic quantum mechanics.

In the present paper, we have considered an isotropic
cutoff in the momentum space
by introducing a single parameter $\lambda$;
$p=|\p| < \lambda$.
We expect that the variety
of the quantum walk models depending on
lattice structure on which a quantum walker exists
and choice of a quantum die represented
by a unitary matrix will be
partially realized by changing how to introduce
cutoff in the theory.
The present isotropic cutoff
is too simple and thus we have some 
differences in symmetry of distribution functions
between the graphs (a) of $\nu^{(d)}$
for the SQW$^{(d)}$ 
and (b) of $\nu^{(d)}_{\rm Dirac}$ 
obtained from the Dirac equations
in Figs. \ref{fig:fig_SatoKatori_4} and
\ref{fig:fig_SatoKatori_5}
in the $d=2$ case.

In the previous paper on the SQW$^{(2)}$ \cite{WKKK08}
an interesting phenomenon called the {\it localization}
of quantum walker at the starting point
was reported (see also \cite{IKK04,MKK07}).
In the present models derived from the Dirac equation,
however, 
the localization phenomenon will not be expected.

In the present study, we have derived the limit
distributions for the $d=1$ and $d=2$
systems from the result for the original
Dirac equation in the three dimensions.
The dependence of the results on the dimensionality $d$
will be summarized as the following,
\begin{equation}
\mu_{\rm Dirac}^{(d)}(v; \lambda)
={\rm const.} \left(\frac{mc}{\lambda}\right)^d
\frac{1}{(1-v^2/c^2)^{(d+2)/2}}
{\bf 1}\left( \frac{v}{c} <
\left\{1+ \left(\frac{mc}{\lambda}\right)^2 \right\}^{-1/2}
\right).
\label{eqn:form1}
\end{equation}
Note that, 
if we define the fifth gamma matrix by
$\gamma_5=\gamma_1 \gamma_2 \gamma_3 \gamma_4$
and consider the Hamiltonian operator of the form
\begin{equation}
\widehat{\cal H}=\gamma_4
\left( \sum_{k=1}^{3} c \hbar \gamma_k \frac{\partial}{\partial x_k}
+ c \hbar \gamma_5 \frac{\partial}{\partial x_5}
+ m c^2 \right),
\label{eqn:H4}
\end{equation}
we can discuss the quantum walk in the four-dimensional space
with the four-component qubit.
Also in this case, 
we can obtain the limit distribution,
which has the form (\ref{eqn:form1})
with $d=4$ as shown in Appendix A.

\begin{acknowledgments}
The present authors would like to thank
E. Segawa for useful comments on the present work.
This work was partially supported by the Grant-in-Aid
for Scientific Research (C) (No.21540397) of
Japan Society for the Promotion of Science.
\end{acknowledgments}

\appendix
\section{Four-Dimensional System}

The diagonalization of the Hamiltonian matrix
corresponding to (\ref{eqn:H4})
can be done in the four-component
momentum space $\p=(p_1, p_2, p_3, p_5)$
as follows,
\begin{equation}
{\cal H}(\p)=U(\p)^{-1}
E(p) \gamma_4 U(\p)
\label{eqn:H4b}
\end{equation}
with
\begin{equation}
U(\p)=\frac{1}{\sqrt{2E(p)}}
\left( \sqrt{E(p) + mc^2} \, I_4
+ \im \sum_{k=1,2,3,5} 
\frac{c \gamma_k p_k}
{\sqrt{E(p)+mc^2}} \right),
\label{eqn:H4c}
\end{equation}
where $E(p)=\sqrt{(p_1^2+p_2^2+p_3^2+p_5^2)c^2+(mc^2)^2}$.
Let $\V(t)=(V_1(t), \dots, V_4(t))$ be the four-component
pseudovelocity.
When we introduce an isotropic cutoff in the model
(\ref{eqn:cutoff}),
the weak limit theorem is obtained as follows;
for any $\alpha_j \in \N_0, 1 \leq j \leq 4$,
\begin{equation}
\lim_{t \to \infty}
\left\langle \prod_{j=1}^{4} V_j(t)^{\alpha_j} 
\right\rangle_{\lambda}
=\prod_{j=1}^4 \int_{-\infty}^{\infty} \frac{d v_j}{c}
\prod_{j=1}^{4} v_j^{\alpha_j}
\, \nu_{\rm Dirac}^{(4)}(\v; \lambda, \q)
\label{eqn:4dimDirac0}
\end{equation}
with
\begin{equation}
\nu_{\rm Dirac}^{(4)}(\v; \lambda, \q)
=\mu_{\rm Dirac}^{(4)}(v; \lambda)
{\cal M}_{\rm Dirac}^{(4)}(\v; \q)
\label{eqn:4dimDirac1}
\end{equation}
where
\begin{eqnarray}
&&\label{eqn:4dimDirac2}
\mu_{\rm Dirac}^{(4)}(v;\lambda)
=\frac{2}{\pi^2}
\left( \frac{mc}{\lambda} \right)^4
\frac{1}{(1-v^2/c^2)^3}
{\bf 1} \left(
v < \sqrt{\frac{\lambda^2 c^2}{(mc)^2+\lambda^2}} \right),\\
\label{eqn:4dimDirac3}
&&
{\cal M}_{\rm Dirac}^{(4)}(\v;\q)
=1+\sum_{j=1}^{4} {\cal M}_{\rm Dirac}^{(4)}(\q) v_j
\label{eqn:4dimDirac4}
\end{eqnarray}
with
\begin{eqnarray}
&&
{\cal M}_{\rm Dirac}^{(4,1)}(\q)
=2 {\rm Re} (q_1\bar{q_4}+q_2\bar{q_3}), \quad
{\cal M}_{\rm Dirac}^{(4,2)}(\q)
=2 {\rm Im}(\bar{q_1} q_4- \bar{q_2} q_3),
\nonumber\\
&&
{\cal M}_{\rm Dirac}^{(4,3)}(\q)
=2 {\rm Re}(q_1\bar{q_3}-q_2\bar{q_4}), \quad 
{\cal M}_{\rm Dirac}^{(4,4)}(\q)
=2 {\rm Im}(\bar{q_1} q_3+ \bar{q_2} q_4). 
\label{eqn:4dimDirac5}
\end{eqnarray}


\begin{thebibliography}{99}
\bibitem{ADZ93}
Y. Aharonov, L. Davidovich, and N. Zagury,
Phys. Rev. A {\bf 48}, 1687 (1993).

\bibitem{Mey96}
D. A. Meyer,
J. Stat. Phys. {\bf 85}, 551 (1996).

\bibitem{NV00}
A. Nayak and A. Vishwanath,
e-print quant-ph/0010117.

\bibitem{ABNVW01}
A. Ambainis, E. Bach, A. Nayak, A. Vishwanath, and
J. Watrous, in
Proc. of the 33rd Annual ACM Symp. on Theory
of Computing
(ACM Press, New York, 2001), pp.37-49.

\bibitem{TM02}
B. C. Travaglione and G. J. Milburn, 
Phys. Rev. A {\bf 65}, 032310 (2002).

\bibitem{Kon02}
N. Konno,
Quantum Inf. Process {\bf 1}, 345
(2002).

\bibitem{Kem03}
J. Kempe,
Contemp. Phys. {\bf 44}, 307 (2003).

\bibitem{Amb03}
A. Ambainis,
Int. J. Quantum Inf. {\bf 1}, 507
(2003).

\bibitem{Ken06}
V. M. Kendon,
Int. J. Quantum Inf. {\bf 4}, 791 (2006).

\bibitem{BBCKPX08}
P. Biane, L. Bouten, F. Cipriani, N. Konno, 
N. Privault, Q. Xu, 
{\it Quantum Potential Theory}, Lecture Notes in Mathematics, 1954
(Springer, Berlin, 2008).

\bibitem{Gro97}
L. K. Grover, Phys. Rev. Lett. {\bf 79}, 325 (1997).

\bibitem{NC00}
M. A. Nielsen and I. Chuang,
{\it Quantum Computation and Quantum Information}
(Cambridge University Press, Cambridge, England, 2000).

\bibitem{Amb04}
A. Ambainis,
in Proc. 45th Annual IEEE Symp. on Foundations of
Computer Science
(Piscataway, NJ, IEEE, 2004), pp.22-31.

\bibitem{OKAA05}
T. Oka, N. Konno, R. Arita, and H. Aoki,
Phys. Rev. Lett. {\bf 94}, 100602 (2005).

\bibitem{KFK05}
M. Katori, S. Fujino, and N. Konno,
Phys. Rev. A {\bf 72}, 012316 (2005).

\bibitem{Str06}
F. W. Strauch,
Phys. Rev. A {\bf 73}, 054302 (2006).

\bibitem{BES07}
A. J. Bracken, D. Ellinas, and I. Smyrnakis,
Phys. Rev. A {\bf 75}, 022322 (2007).

\bibitem{Str07}
F. W. Strauch,
J. Math. Phys. {\bf 48}, 082102 (2007).

\bibitem{Kon05}
N. Konno,
J. Math. Soc. Jpn, {\bf 57}, 1179 (2005).

\bibitem{GJS04}
G. Grimmett, S. Janson, and P. F. Scudo,
Phys. Rev. E {\bf 69}, 026119 (2004).

\bibitem{IKK04}
N. Inui, Y. Konishi, and N. Konno,
Phys. Rev. A {\bf 69}, 052323 (2004).

\bibitem{MKK07}
T. Miyazaki, M. Katori, and N. Konno, Phys. Rev. A {\bf 76}, 
012332 (2007).

\bibitem{WKKK08}
K. Watabe, N. Kobayashi, M. Katori, and N. Konno,
Phys. Rev. A {\bf 77}, 062331 (2008).

\bibitem{SK08}
E. Segawa and N. Konno,
Int. J. of Quantum Inf., 
{\bf 6}, 1231 (2008).

\bibitem{Kon09}
N. Konno, 
Quantum Inf. Process {\bf 8}, 387
(2009).

\bibitem{BD64}
J. D. Bjorken and S. D. Drell,
{\it Relativistic Quantum Mechanics}
(McGraw-Hill, New York, 1964).

\bibitem{Dir58}
P. A. M. Dirac,
{\it The Principles of Quantum Mechanics},
4th edn (Clarendon Press, Oxford, 1958).

\bibitem{IZ80}
C. Itzykson and J-B. Zuber,
{\it Quantum Field Theory}
(McGraw-Hill, New York, 1980).

\bibitem{BFM05}
A. J. Bracken, J. A. Flohr, and G. F. Melloy,
Proc. R. Soc. A {\bf 461}, 3633 (2005).

\bibitem{FW50}
L. L. Foldy and A. Wouthuysen,
Phys. Rev. {\bf 78}, 29 (1950).

\bibitem{Tani51}
S. Tani, 
Prog. Theor. Phys. {\bf 6}, 267 (1951).

\end{thebibliography}

\end{document}